# Defect Detection Efficiency: A Combined approach
Rashmi N, Suma V

## Abstract
*Survival of IT industries depends much upon the development of high quality and customer satisfied software products. Quality however can be viewed from various perspectives such as deployment of the products within estimated resources, constrains and also being defect free. Testing is one of the promising techniques ever since the inception of software in the global market.*

*Though there are several testing techniques existing, the most widely accepted is the conventional scripted testing. Despite of advancement in the technology, achieving defect free deliverables is yet a challenge. This paper therefore aims to enhance the existing testing techniques in order to achieve nearly zero defect products through the combined approach of scripted and exploratory testing. This approach thus enables the testing team to capture maximum defects and thereby reduce the expensive nature of overheads. Further, it leads towards generation of high quality products and assures the continued customer satisfaction.*

## 1. Introduction

The success of any software product depends on several factors such as cost of the product, quality of the product, on time delivery of the product, marketing strategy involved and so on which influences quality. High quality software is one which is error free, produces predictable results with less manageable efforts, understandable, dependable and efficient[9]. However various factors affect quality such as quality of the software process, quality of the people, the quality of the standards in the organization. Therefore, quality can be visualized as a mathematical expression (1) where,

**Rashmi N**, Information Science & Engineering, Dayananda Sagar College of Engineering, Bangalore, India, e-mail: rashmi_jul@yahoo.com

**Suma V**, Information Science & Engineering, Dayananda Sagar College of Engineering, Bangalore, India, e-mail: sumavdsce@gmail.com

$$\text{Software Quality} = \sum_{i=1}^{n} f(\text{People Quality} + \text{Process Quality})$$

(1)
Where, i = 1 requirement phase, n = maintenance phase of software development process [9].

Software Engineering is a systematic approach for developing high quality software. There are two important Quality practices followed in the industry namely quality assurance and quality control. Quality Assurance is the process of assuring quality in the development process through various approaches such as inspections, walkthrough, reviews, audits and assessments, training programs and further through metrics and measurements. Quality Control involves monitoring the software development process to ensure that quality assurance procedures and standards are being followed through testing techniques.

Testing can be manual or automated. There are number of testing techniques used in the industry depending on the type of the software under development. Despite of existence of various testing techniques[2][4] there still prevails a continuing demand from the customers towards test coverage, identifying defects, learning the product, assessing the risks involved with product, performance of the product[12] etc. Hence, delivering high quality software is still a challenge. This is because the number of defects found in software depends mainly on the skill of the tester testing the software. The conventional scripted testing does give little importance to the skills of the tester. It is more of checking the conformance between the requirements specification and the actual behaviour of the software while exploratory testing does not have any predefined test cases as the tester simultaneously learns designs and executes the test cases.

Though automation tools may be used, however, using automation tool for testing all types of software is not possible due to various reasons. In order to address this issue, this research intends to explore the significance and impact of Exploratory testing in addition to Conventional scripted testing practices. Integration of Exploratory testing with the Conventional scripted testing techniques improves the efficiency of the software to a larger extent.

This paper is organized as follows. The first section is the introduction to the combined approach. This describes the significance of the combined approach. Second section is the literature survey. This gives a good understanding of the existing testing techniques. Third section is the description of the combined approach. Fourth section is the research methodology. Fifth section is the case study section. Here the data from the quality department of a company is taken and analyzed. The last section is the inferences section. This gives the observations made on the analysis of the data.

The scope of this paper is to emphasize the importance of exploratory testing in combination with the Conventional scripted testing techniques.

## 2. Literature Review

The purpose of software testing is to provide a framework having a set of disciplines and approaches to test an application which ensures that development process is consistent leading towards the generation of high quality software.

In 2002, C. Andersson et al. in [1] presented a qualitative survey of the verification and validation processes at 11 Swedish companies. The purpose was to exchange the information between the companies. It is concluded from the survey that there are substantial differences between small and large companies. In large companies, the documented process is emphasized while in small companies, single key persons have a dominating impact on the procedures. Large companies use commercial tools while small companies in-house tools or use shareware.

In 2004, Juristo, N et al. in [2] analyzed the maturity level of the knowledge about testing techniques by examining existing empirical studies about these techniques. In 2000, J. A. Whittaker in [3] answers questions from developers how bugs escape from testing. Undetected bugs come from executing untested code, difference of the order of executing, combination of untested input values, and untested operating environment. A four-phase approach was described in answering to the questions. By carefully modeling the software's environment, selecting test scenarios, running and evaluating test scenarios, and measuring testing progress, the author offers testers a structure of the problems they want to solve during each phase.

In 1987, V. R. Basili et al. in [4] apply an experimentation methodology to compare three state-of-the-practice software testing techniques: a) code reading by stepwise abstraction, b) functional testing using equivalence partitioning and boundary value analysis, and c) structural testing using 100 percent statement coverage criteria. The study compares the strategies in three aspects of software testing: fault detection effectiveness, fault detection cost, and classes of faults detected. The major results of this study are the following. 1) With the professional programmers, code reading detected more software faults and had a higher fault detection rate than did functional or structural testing, while functional testing detected more faults than did structural testing, but functional and structural testing was not different in fault detection rate. 2) In one advanced student subject group, code reading and functional testing were not different in faults found, but were both superior to structural testing, while in the other advanced student subject group there was no difference among the techniques. 3) With the advanced student subjects, the three techniques were not different in fault detection rate. 4) Number of faults observed, fault detection rate, and total effort in detection depended on the type of software tested. 5) Code reading detected more interfaces than did the other methods. 6) Functional testing detected more control faults than did the other methods. 7) When asked to estimate the percentage of faults detected, code readers gave the most accurate estimates while functional testers gave the least accurate estimates.

In 2003, A. Tinkham et al. in [5] discuss how exploratory testers differ, what they know, questioning strategies used by an exploratory tester, the role of heuristics during exploratory testing.

In 2000, Bach, J in [6] explains the session based test management, one of the popular exploratory testing styles with an example. Also the author explains the tool support and about the metrics used during session based test management.

In 2007, Juha Itkonen et al. in [7] performed a controlled experiment to compare the defect detection efficiency of exploratory testing (ET) and test case based testing (TCT). Based on the experiment conducted, the authors make few important observations such as lack of benefit in terms of defect detection efficiency of using predefined test cases in comparison to an exploratory testing approach.

In 2005, Juha Itkonen et al. in [8] provide an insight into Exploratory Testing, its applicability, benefits, and shortcomings. Furthermore, they describe how exploratory testing can be utilized across the industries by conducting interviews with testers in seven companies who share their experiences of performing exploratory testing along with sufficient data as a proof for the same.

In 2011, T. R. Gopalakrishnan et al. in [9] provide an empirical investigation of several projects through a case study comprising of four software companies having various production capabilities. The aim of this investigation was to analyze the efficiency of test team during software development process. The study indicates very low-test efficiency at requirements analysis phase and even lesser test efficiency at design phase of software development. Subsequently, the study calls for a strong need to improve testing approaches using techniques such as dynamic testing of design solutions in lieu of static testing of design document.

In 2008, V. Suma et al. in [10] provides information on various methods and practices supporting defect detection and prevention leading to thriving software generation. Inspection was considered as one of the best defect prevention techniques which can reduce test effort since most of the static defects could be captured through inspection. Authors therefore recommend implementation of inspections to reduce high expenses related to testing.

Authors of [12] perform a controlled experiment to show the significance of exploratory testing. Through the experiment they prove that exploratory testing is as effective as scripted testing.

## 3. Combined Approach

Despite of existence of several approaches to enhance testing efficiency, the defect removal efficiency has not improved beyond 85% [11]. This paper therefore aims to validate the need for combined approach of exploratory testing and the conventional scripted testing approach.

In scripted testing the tests are designed and documented during the early stages of software development but are executed at later stages by a different tester. In exploratory testing tests are designed and executed at the same time but are not documented. According to a survey [11] there are several reasons behind accepting the scripted testing as a default practice across companies. The reasons include advancement in technology, rise in costs, increase in the size of the application etc. While scripted testing emphasizes accountability and decidability of tests, exploratory testing emphasizes adaptability and learnability of the software under test [5].

Exploratory testing is a novel approach of testing where the tester actively controls the design of the tests as those tests are performed and uses information gained while testing to design new and better tests. There are no predefined test cases in exploratory testing. No test case document is maintained before we could execute the test cases. It is an approach to testing where the learning, test design and test execution happens in a simultaneous manner instead of designing the tests in the early stages of software development [5] [6].

Thus, exploratory testing provides unique opportunities for exploring the potential benefits of the experience and the knowledge of the tester. The knowledge of the tester can however be viewed from two perspectives namely the knowledge acquired by the tester on the application that the tester is testing or may be the knowledge of the platform on which the application is running.

Nevertheless the strength of Exploratory Testing, it is not a replacement for the conventional scripted testing. However combined approach of scripted and exploratory testing would be useful in certain situations [8].

Hence, use of structured methods and initiatives can improve the efficiency of exploratory testing [11] when carefully combined with mainstream scripted execution. Also the combined approach would result in the increased effectiveness of the testing function.

Fig.1 below describes the significance of the combined approach. In the figure first the module is tested using the conventional scripted testing. The outcome of the scripted testing is named as the tested module. The tested module when undergoes exploratory testing it results in the fine module. This fine module is more defect free than the tested module.

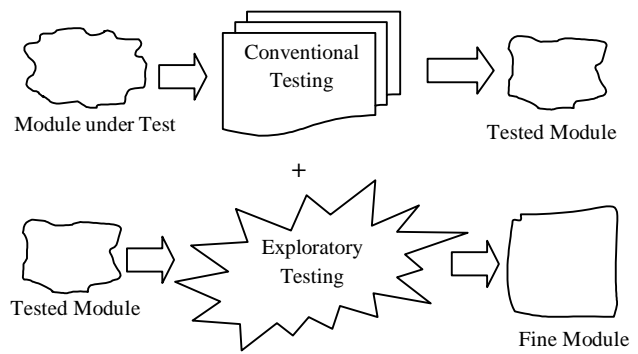

**Figure 1: The Combined Approach**

## 4. Research Methodology

The aim of this research paper is to emphasize the importance of the combined approach. This research therefore focuses on the defect detection efficiency of both the approaches with a case study comprising of an investigation carried out in one of the leading software company with CMMI level 5 certification. The company is a product based company providing digital printing solutions. There are four projects in the case study. All projects are non-critical in nature and of embedded domain type. The projects are implemented using programming languages such as C, C++ on platforms such as WINDOWS and LINUX. All projects are developed using V-model and are of standalone type. All projects are of maintenance type projects.

The sample data presented in this paper comprises of small category of projects which can be developed within 1000 person hours.

Data collection is through the data centers and quality assurance departments of the above mentioned company. Data analysis is done using comparison of testing techniques using project success as a criteria measured through defect capturing capability.

From the data analysis carried out through the comparative study of techniques indicates that combined approach is better than isolated approach.

## 5. Case Study

The objective of this paper is to emphasize the benefit obtained by performing exploratory testing in addition to conventional scripted testing. Hence, this paper presents a case study where an investigation of several projects is carried in a leading product based software industry [7] [8] [9].

Table 1 illustrates the data collected from the company. It depicts the defect capturing capability of the testing team practicing scripted testing. The table provides information on the total development time required for the project completion which is measured in terms of person hours, the choice of process model followed, number of testers assigned in addition to the time scheduled for testing. The table further specifies the number of defects estimated and captured by the testing team.

**Table 1: Scripted Approach**

| Sl.No | Parameters | P1 | P2 | P3 | P4 |
|---|---|---|---|---|---|
| 1 | Project Development Time(*) | 324 | 243 | 432 | 162 |
| 2 | Development life cycle model | V-Model | V-Model | V-Model | V-Model |
| 3 | No. of Testers | 6 | 9 | 4 | 3 |
| 4 | Scripted Test time of the project (*) | 270 | 162 | 360 | 135 |
| 5 | No. of defects captured by testing team | 451 | 251 | 972 | 1022 |

(*) – Person Hours. The Project Development Time is expressed in person hours which are given by Project Development Time = (9 hours of work per day)*(number of personnel)*(number of months required)

Table 2 illustrates the defect capturing capability of the testing team practicing the combined approach. The parameters considered here are the project development time, testing time, number of defects captured, number of testers and so on.

**Table 2: Combined Approach**

| Sl. No | Parameters | P1 | P2 | P3 | P4 |
|---|---|---|---|---|---|
| 1 | Project Development Time (*) | 324 | 243 | 432 | 162 |
| 2 | Development life cycle model | V-Model | V-Model | V-Model | V-Model |
| 3 | No. of Testers | 6 | 9 | 4 | 3 |
| 4 | Test time of the | 270 | 162 | 360 | 135 |
| 5 | ET test time(*) | 27 | 16.2 | 36 | 13.5 |
| 6 | No. of defects captured by testing team | 602 | 851 | 1244 | 1644 |

There are four projects being studied in the case study [9] [10]. The same projects were compared using the scripted and the combined approach. There are five parameters being used in the tables Table 1 and Table 2. They are defined as follows. First parameter is the total development time of the project which includes the coding time and the testing time of the project, second is the type of the development model used, third is the total number of testers in the project. The testers are having the relevant experience in their respective domain. Fourth parameter is the amount of time allotted for testing the product. This involves the time spent on unit, in the combined approach 10-15% of the total test time is allotted for exploratory testing. We have considered 10% exploratory test time in this case study.

## 6. Inferences

When we compare the scripted and the combined approaches we find the combined approach yielding more number of defects in the same time as the time allotted for scripted testing. This is shown in the form of a table, Table 3.

**Table 3: Comparison Chart**

| Projects | Scripted Approach | Combined Approach | Additional defects detected using combined |
|---|---|---|---|
| P1 | 451 | 602 | 151 |
| P2 | 251 | 851 | 600 |
| P3 | 972 | 1244 | 272 |
| P4 | 1022 | 1644 | 622 |

This can also be represented graphically as shown below.

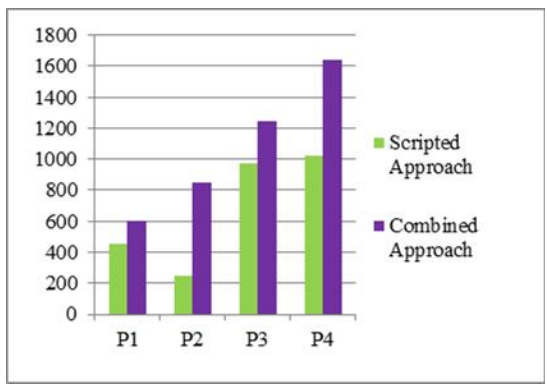

**Figure 2: Comparison Graph**

Hence from the above analysis we infer that the combined approach is more efficient in terms of the number of defects found and in terms of the time spent on testing. The only concern here is that the testers performing the exploratory testing are having relevant domain experience.

However, this inference is made on projects which are developed within 1000 person hours.

Further, this paper provides details of defects captured in small complex projects. It needs further investigations to be made on various other defect influencing parameters in addition to defect estimation and defect capturing capability of combined approach on varied complex projects.

## 7. Conclusions

Quality of a product is some value to a person. Different customers will perceive having the same product having different levels of quality. Hence delivering high quality software is always a great challenge to software companies. Defect free product is deemed to be one of the needs for achieving high quality products. Software testing is the last opportunity for the testers to detect as many defects as possible before the product is delivered to the customers.

Despite of existence of a spectrum of testing approaches, techniques and tools, each of these approaches or the techniques and tools has their own strengths and weakness. Therefore, identifying the appropriate combination of these approaches or techniques is a challenging task. One such combination is the combined approach which is discussed in this paper. Though scripted and exploratory testing techniques individually contribute greatly towards the quality of the product when combined they increase the defect detection efficiency to a greater extent. Hence, the quality of the product is even more improved.

### Acknowledgment

Authors would like to thank all the industry people who extended their valuable support and help in compliance within the framework of non-disclosure agreement.

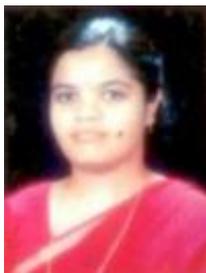
**Rashmi N. B.E, M.Tech, (Ph.D)**
Currently working as an Assistant Professor in Dept. of Information Science & Engineering at Dayanand Sagar College of Engineering. Has 7+ years of industrial experience in Software Testing and also in acedemics. Currently, she is pursuing Ph.D under the guidance of Dr.Suma V under Visweswaraya Technological University.

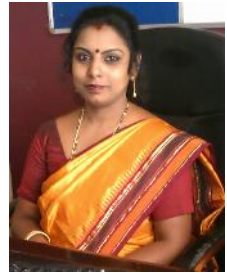
**Dr. Suma V, B.E., M.S, Ph.D**
Dr. Suma is currently the Dean, Research and Industry Incubation Center, Dayananda Sagar Institutions and having 16+ years of experience. She has more than 80 International publications including IEEE, ACM, ASQ, Crosstalk, IET Software, Inderscience publishers, Springer, NASA, UNI trier, Microsoft, CERN portals and obtained several best paper awards. She is an invited author for International Book Chapter and her works are given as projects at various universities. She is an International Reviewer, TCP and Advisory Board Member, key note speaker for several International Conferences.